\documentclass[aps,twocolumn,showpacs,showkeys,groupedaddress]{revtex4}

\usepackage{epsfig}
\usepackage{epsf}

\def\slash#1{\setbox0=\hbox{$#1$}
   \dimen0=\wd0 \setbox1=\hbox{/} \dimen1=\wd1
   \ifdim\dimen0>\dimen1 \rlap{\hbox to \dimen0{\hfil/\hfil}} #1
   \else  \rlap{\hbox to \dimen1{\hfil$#1$\hfil}} / \fi}
   
\def\intp{\int \frac{d^4 p}{(2\pi)^4}}

\def\intw{\int d\omega \rho(\omega)}

\begin{document}

\title{Axial Vector Coupling and Chiral Anomaly in the Spectral Quark Model\footnote{This work is dedicated to Lincoln Almir Amarante Ribeiro (in memorian).}}%

\author{L. C. Ferreira}
\author{A. L. Mota}
\email{motaal@ufsj.edu.br}
\affiliation{Departamento de Ci\^{e}ncias Naturais, Universidade Federal de S\~{a}o Jo\~{a}o del Rei,
C.P. 110,  CEP 36301-160, S\~ao Jo\~ao del Rei, Brazil}

\begin{abstract}
We studied the Adler-Bardeen-Bell-Jackiw anomaly in the context of
a finite chiral quark model known as the Spectral Quark Model.
Within this model, we obtain the general non-local form of the axial vertex
compatible with a non vanishing axial coupling, in the chiral limit. 
The triangle anomaly is computed and we show that the obtained dependence of the axial 
vertex with the spectral mass is necessary to ensure both finiteness and
the correct violation of the chiral Ward-Takahashi identity.
\end{abstract}

\maketitle

\section{Introduction}

The non-perturbative low energy behavior of Quantum Chromodynamics (QCD) is well described by a series of effective 
models \cite{LSM,Nambu:1961fr,Nambu:1961tp} in both zero and finite temperature and densities 
\cite{RamMohan:1976fu,Asakawa:1989bq,Klimt:1990ws,Kunihiro:1991hp,Caldas:2003hj,Caldas:2001wx,Caldas:2000ic,Caldas:1997dw,Andersen:2004ae,Andersen:2008qk,
Petropoulos:1998gt,Roh:1996ek,Goldberg:1983ju,Lenaghan:2000ey,Baym:1977qb,Reinhardt:1987da,Christov:1991se,Ebert:1992ag,Schwarz:1999dj}. 
It is assumed that these models could result from the suppression of the high energy degrees of freedom of QCD (such as gluons),
and a scale that defines the validity of the model has to be introduced. 
In some of these models, the quarks remain as the only degrees of freedom and, as these chiral quark models are usually 
non-renormalizable, in contrast with QCD itself and with mesonic models, these scales (or cut-off) are kept finite throughout the calculations. 
As a consequence, a series of problems \cite{Blin:1987hw,RuizArriola:2002wr} emerge as a reflex of this process. Nevertheless, the light hadrons 
phenomenology is successfully 
described by almost all these models, that, together with the non perturbative nature of QCD at this scale, justifies their employment. 

Some of the results that are jeopardized by the introduction of a cut-off scale are the anomaly dependent results. In particular,
the anomalous transition form factor $F_{\gamma^{*}\pi^{0}\gamma}(Q^2)$, for the process $\gamma^{*} \rightarrow \pi^{0}\gamma$,
can only be correctly reproduced in the limit $Q^2=0$ when no regulator is introduced \cite{Blin:1987hw}. In contrast, 
a finite regularization is necessary to keep the models finite, and it is also necessary to reproduce the expected QCD
factoration at large momenta, $Q^2 F_{\gamma^{*}\pi^{0}\gamma}(Q^2) \rightarrow 2 f_{\pi}$ \cite{RuizArriola:2002wr}, where $f_{\pi}$
is the pion weak decay constant. 

Perturbativelly, the anomaly appears as an ambiguity, represented by surface terms, due to the infinities of the perturbative
calculations. Regularization of the Feynman integrals fixes, {\it a priori}, the result of the computation of these surface
terms, but the undetermined nature of the ambiguity appears as different results for different regularizations. If the undeterminacy
of these terms is kept up to the end of the calculation, as occurs in some recent regularization schemes \cite{Battistel:1998sz,BaetaScarpelli:2001ix,Freedman:1991tk},
it can be shown that the simultaneous transversality for massless fermions in the axial and vector channels is broken
\cite{Jackiw:1999qq}. On one hand, it is possible to fix the ambiguity in order to ensure the conservation of the electromagnetic
Ward-Takahashi identity (violating the chiral Ward-Takahashi identity), as required in QCD in order to explain the
anomalous decays. On the other hand, it is also possible to fix the ambiguity in order to satisfy the chiral Ward-Takahashi identity
and to violate the electromagnetic one, as in t'Hoofts proton decay calculation \cite{tHooft:1976up}.

This undetermined character of the anomaly \cite{Jackiw:2000dh} and the implications of the presence of ambiguous
terms in Quantum Field Theory calculations are in the heart of some recent controversies in the study of
Lorentz and CPT violations in QED \cite{Jackiw:1999yp,PerezVictoria:1999uh,Chung:1999pt,Andrianov:2001zj,Bonneau:2000ai,Bonneau:2006ma}. 
The picture in chiral quark models is somewhat different.
Their divergent non-renormalizable character implies in a dependence with an specific regularization scheme, which, 
{\it a priori}, fixes the ambiguous integrals. So, one cannot expect, in these models, the undetermined character of
the anomaly to manifest itself. The regularization schemes employed usually fix the vector gauge symmetry, and
the transversality of the vector currents is guaranteed from the very beginning, reproducing for this reason the expected 
final result.

A recent chiral quark model, namely, the Spectral Quark Model (SQM) \cite{Enrique,RuizArriola:2003wi,Megias:2004uj,Arriola:2006ds,Broniowski:2007fs}, 
has some interesting features that can
be explored to study the presence of the undetermined character of the anomaly, as well as to correct some of the fails 
on chiral quark models. For example, it was shown \cite{Enrique} that the SQM solves
the conflict between the anomaly normalization condition, $F_{\gamma^{*}\pi^{0}\gamma}(0)=\frac{1}{4\pi f_\pi}$, and the 
factorization at large momenta $Q^2 F_{\gamma^{*}\pi^{0}\gamma}(Q^2) \rightarrow 2 f_{\pi}$. The Spectral Quark Model
is based on the Lehmann representation for the quark propagator \cite{Itzykson:1980rh}, and in the solutions for the chiral
and electromagnetic Ward-Takahashi identities using the gauge technique \cite{Delbourgo:1977jc,Delbourgo:1978bu}, 
resulting in a finite quark model.
Being finite, one can speculate (i) if the model correctly reproduces the anomalous results, (ii) if the 
freedom on the choice of the Ward-Takahashi identity to be violated by the anomaly is still present, as discussed above,
and under which conditions one can obtain the expected violation of the chiral Ward-Takahashi identity and conservation of
the electromagnetic ones in QCD.

Question (i) is already positively answered in Ref. \cite{Enrique}. An important ingredient to this answer is the
spectral version of the Goldberger-Treiman relation \cite{Goldberger:1958tr}, a fact that we will stress later. In this
paper, we intend to explore the answers to question (ii). As we shall see, the specific form of the axial
vector coupling plays an important role in this issue. 

Non-local axial vector vertex for constituent quarks allows the presence of an axial vector coupling constant $g_A$, and
leading order $N_c$ effects can result in $g_A \neq 1$ \cite{Broniowski:1993nb}, even in the large $N_c$ limit. In contrast with the
pseudo-scalar pion quark coupling constant $g_{\pi qq}$, which vanishes in the chiral limit \cite{Goldberger:1958tr}, one should
not expect the axial coupling constant to vanish at this limit - model estimatives to $g_A$ lie in
the range $0.4 < g_A < 0.9$ \cite{Carter:2002xx}, a result that is compatible with the axial vector coupling constant of the nucleon, 
$G_A/G_V=1.2670 \pm 0.0035$, deduced from
neutral beta decay measurements \cite{Groom:2000in} and from the non-relativistic relation $G_A=\frac{5}{3}g_A$. In the context
of the SQM, a dependence of $g_A$ with positive powers of the spectral mass $\omega$, as occurs for the pion quark
coupling constant via the Goldberger-Treiman relation, would be desirable, since positive momenta of the spectral
distribution $\rho(\omega)$ guarantees the finiteness of the amplitudes. We will show that a spectral $g_A(\omega)$
with these characteristics is compatible with a non-vanishing axial coupling in the chiral limit.


In this paper, we: (i) discuss the role of the spectral version of the Goldberger-Treiman relation in the pseudoscalar two point function, showing its
importance to the finiteness of this amplitude and to the obtaining of the Nambu-Goldstone mode; (ii) obtain a more general expression
for the axial vector vertex in the context of the Spectral Quark Model, which includes the possibility of a non unitary axial
coupling and (iii) by applying the gauge technique, obtain the probability amplitude for the axial-vector-vector process 
in the SQM, including non unitary axial coupling, and show that a dependence on the spectral mass for the axial coupling can
generate an ambiguity free result, preserving the vector Ward identity and violating the axial one, as expected for QCD.

This paper is organized as follows: in section II we briefly review the spectral quark model and some of its consequences.
In section III, we analyze the role of the Goldberger and Treiman relation in the finiteness of the pseudoscalar two point function.
In section IV we present the construction of a general form of the axial vertex in the SQM compatible with a
non vanishing axial coupling in the chiral limit. In section V, we compute the axial-vector-vector amplitude by employing
the axial vertex obtained in section IV. We discuss the violation of the chiral Ward-Takahashi identity, and the role of
the axial vertex on this result. Finally, in section VI we present the conclusions.

\section{The Model}

The Spectral Quark Model is based on the introduction of the generalized Lehmann representation for the quark propagator
\begin{equation}
S(p)=\int_C d\omega \frac{\rho(\omega)}{{\slash p}-\omega}, \label{Sp}
\end{equation}
where $\omega$ is the spectral mass of the quark and
$\rho(\omega)$ is the spectral distribution, that acts as a regulator. The $\omega$ integral is supposed
to be valuated in a suitable complex contour $C$ (suppressed in our notation, from now on). 

The spectral function $\rho(\omega)$ needs not to be completely determined, although it is possible to find an
explicit form to it if the vector-meson dominance of the pion form factor is assumed \cite{Enrique,Volmer:2000ek}. 
For the reproduction of most of the
light mesons phenomenology, it is sufficient to know some of the moments of the quark spectral function,
determined via physical conditions such as normalization of the quark propagator, which implies in
\begin{equation}
\rho_0=\intw = 1,
\end{equation}
and finiteness of hadronic observables, implying in
\begin{equation}
\rho_n=\intw \omega^n = 0, \label{positivemoments}
\end{equation}
for $n=1,2,3$ or $4$. Physical observables are proportional to the negative moments
\begin{equation}
\rho_{-n}=\intw \frac{1}{\omega^n} 
\end{equation}
or to the logarithmic moments
\begin{equation}
\rho^{\prime}_{n}=\intw \omega^n \log(\omega^2) .
\end{equation}

Finiteness is guaranteed by the vanishing of the positive moments, Eq. (\ref{positivemoments}). As a consequence, 
negative and log moments can be fixed from the finite values of hadronic observables such as the quark condensate
for one flavor
\begin{equation}
<\bar{q}q> = -\frac{N_c}{4 \pi^2}\rho^{\prime}_3,
\end{equation}
the vacuum energy density
\begin{equation}
B=-\frac{3N_c}{16 \pi^2}\rho^{\prime}_4,
\end{equation}
and so on. 
The quark propagator (\ref{Sp}) can be parameterized as 
\begin{equation}
S(p)=Z(p) \frac{{\slash p}+m(p)}{p^2-m^2(p)},
\end{equation}
with the mass function given by
\begin{equation}
m(p)=\frac{\int d\omega \frac{\omega \rho(\omega)}{p^2-\omega^2}}{\int d\omega \frac{\rho(\omega)}{p^2-\omega^2}} \label{massfunction}
\end{equation}
and the wave function renormalization by
\begin{equation}
Z(p)=(p^2-m^2(p)) \int d\omega \frac{\rho(\omega)}{p^2-\omega^2}.
\end{equation}
In what follows, we will refer to the mass function at $p^2=0$ as $m=m(0)$. The results presented here will not
depend explicitly on $m$, but it will be necessary in order to make contact with the standard representation
of the partial conservation of the axial current (PCAC).

The detailed determination of the spectral function moments, as well as the development of
the SQM to the low-energy hadron phenomenology is presented in \cite{Enrique}, and is not the aim of the present
contribution.

To proceed with the computation of N-point functions in the SQM, the vertex functions are defined as particular solutions
of the relevant Ward-Takahashi identities for unamputed Green functions, obtained by applying the gauge technique 
\cite{Delbourgo:1978bu}. This allows the obtaining of linear solutions, whereas the use of amputed Green functions would imply
in the appearance of non-linear solutions to the Ward-Takahashi identities. The vector Ward-Takahashi identity (VWI), for the vector-quark-quark
vertex $\Lambda^{\mu a}_V$, reads
\begin{equation}
(p^{\prime} - p)_{\mu} \Lambda^{\mu a}_V(p^{\prime},p) = S(p^{\prime})\frac{\lambda^a}{2} - \frac{\lambda^a}{2} S(p), \label{WIV}
\end{equation}
where $S(p)$ is given by Eq.(\ref{Sp}) 
and $\lambda_a$ are the Pauli matrices (we are assuming a SU(2) flavor symmetry). A solution to Eq. (\ref{WIV}),
up to transverse pieces, is
\begin{equation}
\Lambda^{\mu a}_V(p^{\prime},p) = \intw \frac{i}{\slash{p}^{\prime}-\omega} \gamma^{\mu} \frac{\lambda^a}{2} \frac{i}{\slash{p}-\omega}.
\end{equation}

The axial vector Ward identity (AWI) reads:
\begin{equation}
(p' - p)_{\mu}\Lambda_{5}^{\mu a}(p', p) = 
S(p')\frac{\lambda^{a}}{2}\gamma_{5} + \gamma_{5}\frac{\lambda^{a}}{2}S(p),\label{WIA}
\end{equation}
and one possible solution to Eq.(\ref{WIA}) for the axial vector to quarks coupling is
\begin{equation}
\Lambda_{5}^{\mu a}(p', p) = \intw \frac{i}{\slash{p}^\prime - \omega} 
\left( \gamma^{\mu} - \frac{2 \omega q^{\mu}}{q^{2}} \right) 
 \frac{\gamma_{5}\lambda^{a}}{2} \frac{i}{\slash{p} - \omega}, \label{AxialVertex1}
\end{equation}
with $q=p^\prime-p$. One can identify the pion pole in (\ref{AxialVertex1}), the dominant term as $q^2 \rightarrow 0$. 
For massless quarks, it cannot be done with
a non spectral propagator, since in this case we have $\rho(\omega)=\delta(\omega)$, and the consequent vanishing of the pole term.

\section{The pseudoscalar two point function}

The pion to quarks coupling can be obtained from the axial vector vertex near the pion pole by using
\begin{equation}
\Lambda_{5}^{\mu a}(p', p) \bigg\arrowvert_{q \rightarrow 0} = - 2 f_\pi \frac{q^\mu}{q^2}\Lambda^a_\pi(p',p),
\end{equation}
resulting in
\begin{equation}
\Lambda^a_\pi(p',p) = \intw \frac{i}{\slash{p}'-\omega}\gamma_{5} \frac{\omega}{f_\pi} \frac{\lambda^{a}}{2} \frac{i}{\slash{p} - \omega}. \label{pionVertex}
\end{equation}

An important consequence of the appearance of the pion pole in solution (\ref{AxialVertex1}) is the obtaining of the 
spectral version of the Goldberger-Treiman relation $g_\pi(\omega)=\frac{\omega}{f_\pi}$. By closing the quark line in 
the pion vertex, Eq.(\ref{pionVertex}), we can obtain the pseudoscalar two-point function
\begin{equation}
\Pi_{PS}(q)=\intw \int_p Tr \Big\{ \frac{i}{\slash{p}+\slash{q}-\omega}\gamma_{5} \frac{\omega}{f_\pi} \frac{\lambda^{a}}{2} \frac{i}{\slash{p} - \omega} \gamma_{5} \Big\}, \label{PiPS}
\end{equation}
where $\int_p$ stands for $\intp$. 
Due to the
spectral conditions, Eq.(\ref{positivemoments}), the divergent terms arising from the computation of Eq.(\ref{PiPS}) vanishes, 
and the final result is finite. In the intermediary calculation, however, an auxiliary regularization scheme is necessary in order
to compute the loop momentum integral before the evaluation of the integral over the spectral mass $\omega$. The use of a 
gauge invariant regularization scheme would enforce the {\it a priori} preservation of the gauge Ward-Takahashi identities. It is
interesting, however, to explore how the correct violation of the AWI in QCD can be obtained in a regularization independent
way. We thus choose to work with the sharp-cutoff regularization scheme, a scheme that violates the VWI, as is widely known.
 
In fact, employing the covariant sharp cutoff regularization scheme to compute the $p$ integral in (\ref{PiPS}), one obtains
\begin{eqnarray}
 & & \Pi_{PS}(q)= \frac{-i}{4\pi^2} \int d\omega \rho(\omega) \frac{\omega}{f_\pi}  \Big\{ 2\Lambda^2 - 2\omega^2 ln\Big( \frac{\Lambda^2}{\omega^2} \Big) + \nonumber \\ 
 & & q^2 \Big( 1 - ln\Big(\frac{\Lambda^2}{\omega^2}\Big) \Big)+  \nonumber \\
 & & 2 q^2 \Big( -1 + \sqrt{1-\frac{4 \omega^2}{q^2}} tanh^{-1}\Big(\frac{1}{\sqrt{1-\frac{4 \omega^2}{q^2}}}\Big) \Big) \Big\}. \label{PIPScut}
\end{eqnarray}
From the first term on the right hand side of Eq.(\ref{PIPScut}) one can clearly see that the Goldberger-Treiman relation is
important in order to make the psedoscalar two-point function finite. If the coupling between the pion and the quarks
was not dependent on $\omega$, this term would be divergent in the limit $\Lambda \rightarrow \infty$.

After applying the spectral conditions (\ref{positivemoments}) on Eq.(\ref{PIPScut}), we get
\begin{eqnarray}
 & & \Pi_{PS}(q)= \frac{-i}{4\pi^2} \int d\omega \rho(\omega) \frac{\omega}{f_\pi}  \Big\{ (q^2 + 2\omega^2) ln(\omega^2) +  \nonumber \\
 & & 2 q^2 \sqrt{1-\frac{4 \omega^2}{q^2}} tanh^{-1}\Big(\frac{1}{\sqrt{1-\frac{4 \omega^2}{q^2}}} \Big) \Big\}. \label{PiPS2}
\end{eqnarray}

From now on, all momentum integrals will be computed employing the covariant sharp cutoff regularization scheme.
The use of this regularization to compute the pseudoscalar two point function introduces surface terms 
that could, in principle, break the Nambu-Goldstone mode \cite{Klevansky:1992qe}. It also generates dependence on the arbitrary
choice of the momentum routing in the loop. In a renormalizable theory it will be no problem: regularization and
symmetries fix these ambiguities \cite{Bonneau:2000ai,Dias:2005tb,Hiller:2005uw}. This is not the case for purely fermionic chiral quark models.
It is interesting, however, to observe how the spectral
regularization corrects this fail: the surface terms that emerges from this computation are 
\begin{eqnarray}
 & & 2q_{\mu}q_{\nu} \int d\omega \rho(\omega) \frac{\omega}{f_\pi} \int_p \Big( \frac{g^{\mu\nu}}{(p^2-\omega^2)^2}-4\frac{p^{\mu}p^{\nu}}{(p^2-\omega^2)^3} \Big) =\nonumber \\
 & & 2q_{\mu}q_{\nu} \int d\omega \rho(\omega) \frac{\omega}{f_\pi} g^{\mu\nu} \omega^2 \frac{i} {2(4\pi)^2\omega^2} = 0
\label{surfacePiPS}
\end{eqnarray}
The details on the computation of surface terms will be presented
on section V, in the context of the chiral anomaly. As we can see, the spectral condition (\ref{positivemoments}) guarantees that Eq.(\ref{surfacePiPS}) vanishes,
since it depends on $\rho_1=\int d\omega \omega \rho(\omega) = 0$. The role played by the Goldberger-Treiman coupling 
$\omega/f_\pi$ becomes clear - if it was not present, then the pseudoscalar two point function, Eq.(\ref{PiPS2}), would be divergent and 
dependent of the ambiguous
result of the momentum integral in the left side of Eq.(\ref{surfacePiPS}). This feature - the dependence of the coupling
with the spectral mass {\it via} the Goldberger-Treiman relation - suggests that a similar dependence on the spectral mass
for the axial vector coupling could be important in the study of the chiral anomaly in the context of the Spectral Quark Model.

\section{The Axial Vector Coupling in SQM}

Eq.(\ref{AxialVertex1}) is one of the possible solutions to the AWI, Eq.(\ref{WIA}). Its
functional form suggests that, for this ansatz, the axial coupling $g_A$ is unitary. As mentioned before, there are
mechanisms that generate contributions to $1-g_A$ of order $N_c^0$, such as the $\pi-A_1$ mixing mechanism \cite{Broniowski:1993nb},
present in chiral models such as the $\sigma$-model\cite{LSM} and the NJL model \cite{Nambu:1961fr}. A general form to
the axial vertex that allows $g_A \ne 1$, including pseudoscalar and pseudovector pion couplings is 
\footnote{The axial-vector and vector vertices could also include terms proportional to $[\gamma^{\mu}, \gamma^{\nu}]$, not included in the present approach.}

\begin{eqnarray}
\Lambda_{A}^{\mu a}(p', p)=\intw \frac{i}{\slash{p}^\prime - \omega} \Big( g_A(q^2,\omega) \gamma^{\mu}  \label{AxialVertex2}  \\
+ h_A(q^2,\omega) q^{\mu}  + f_A(q^2,\omega) q^{\mu} \slash{q} \Big) 
\gamma_5 \frac{\lambda^{a}}{2} \frac{i}{\slash{p} - \omega}, \nonumber 
\end{eqnarray}
where we introduced the spectral form factors $g_A(q^2,\omega)$, $h_A(q^2,\omega)$ and $f_A(q^2,\omega)$. One can recognize
that $g_A(q^2,\omega)=1$, $h_A(q^2,\omega)=-2\omega/q^2$ and $f_A(q^2,\omega)=0$ in the ansatz (\ref{AxialVertex1}).
The evaluation of (\ref{WIA}) with (\ref{AxialVertex2}) gives
\begin{equation}
h_A(q^2,\omega)=-2\omega/q^2 \label{ha}
\end{equation}
and
\begin{equation}
f_A(q^2,\omega)=\frac{1-g_A(q^2,\omega)}{q^2}. \label{fa}
\end{equation}

In what follows we will assume that the axial coupling depends on the spectral mass, but not on the exchanged momenta, 
i.e., $g_A(q^2,\omega)=g_A(\omega)$. One can recognize the poles associated to the Goldstone pion
in the second and third terms on the right hand side of Eq.(\ref{AxialVertex2}) whereas the first term is associated to
the axial vector coupling. Let us denote this term as $\Lambda^{\mu a}_{g_A}(p',p)$, with 
\begin{equation}
\Lambda^{\mu a}_{g_A}(p',p)=\intw g_A(\omega) \frac{i}{\slash{p}'-\omega} \gamma^{\mu}\gamma^5 \frac{\tau_a}{2}\frac{i}{\slash{p}'-\omega}
\end{equation}

For on-shell massless quarks, Dirac equation implies $\slash{p}'=\slash{p}=0$. Thus, we get for the axial vector coupling
with on-shell quarks
\begin{equation}
\Lambda^{\mu a}_{g_A}(p',p)\Big|_{\text{on-shell}}=- \Big\{ \intw \frac{g_A(\omega)}{\omega^2} \Big\} \gamma^{\mu} \gamma^5 \frac{\tau_a}{2} \label{acos}
\end{equation}

Eq.(\ref{acos}) gives us some insight about the functional form of $g_A(\omega)$: if it was proportional to
any positive integer power of the spectral mass $\omega$ greater than $2$ the axial vector coupling for on-shell quarks would be zero.
Yet, assuming $g_A(\omega)=\alpha \omega^n$ with $n \le 2$ and $\alpha$ an arbitrary constant, we obtain a
non vanishing axial coupling in this limit. 
From our previous analysis of the pseudoscalar two-point function, a coupling
constant proportional to an integer positive power of $\omega$ (i.e., $n=1$ or $n=2$) would be desirable in order to render 
finite some amplitudes, and also to avoid regularization ambiguities. In the next section we will see that this feature
is also necessary in the computation of the chiral anomaly in SQM, in order to make it free from ambiguities.

\section{The Chiral Anomaly}

The mechanism of the anomalous symmetry breakdown was co-discovered by Bell and Jackiw \cite{Bell:1969ts} and Adler \cite{Adler:1969gk}.
This violation is related to a probability amplitude that cannot satisfy, simultaneously, the gauge and chiral symmetries.
Nevertheless, which symmetry is violated is a model dependent result: in the anomalous pion decay, the VWI
are to be conserved and the AWI is violated, whereas in the t'Hooft calculation of the proton decay \cite{tHooft:1976up},
the situation is opposite - the global gauge symmetry is violated and the chiral symmetry is preserved.

\begin{figure}[t]
\includegraphics[height=1.4in]{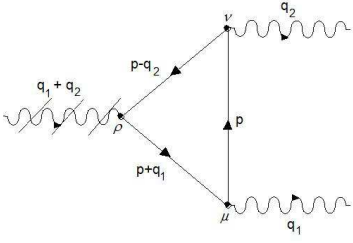}
\caption{The axial-vector-vector 3-point function. Wavy lines correspond to vector currents and wavy slashed line corresponds to the axial vector current.}
\label{fig1}
\end{figure}

The probability amplitude for the axial-vector-vector process is represented by the triangle diagram, depicted on fig. \ref{fig1}.
In order to compute it in the SQM, one needs to find a spectral representation to the vertex function with one axial and one vector
current, $\Lambda^{\mu a;\nu b}_{AV}$, that fulfills the $SU(2)$ vector and axial vector Ward identities. 

For non-local axial vertex in the chiral limit, the axial Ward identity reads:
\begin{eqnarray}
& & -i q_{\mu} \Lambda^{\mu a;\nu b}_{AV}(p',q',p,q) = i \epsilon_{bac} \Lambda^{\nu c}_{A}(p^{\prime},p)\Big|_{g_A=1} \nonumber\\
& & + \Lambda^{\nu b}_{V}(p^{\prime},p+q)\gamma^5 \frac{\tau_a}{2} + \gamma^5\frac{\tau_a}{2} \Lambda^{\nu b}_{V}(p^{\prime}-q,p), \label{LAVAWI}
\end{eqnarray}

The first term on the right hand side of Eq.(\ref{LAVAWI}) is the axial vertex computed with $g_A=1$. This term express the fact that the contraction
of the axial vertex with the exchanged momentum does not depends on $g_A$, as can be easily checked by evaluating $q_{\mu}\Lambda^{\mu a}_A$ from Eq. (\ref{AxialVertex2})
with the use of Eqs.(\ref{ha}) and (\ref{fa}).

The vector Ward identity reads:
\begin{eqnarray}
& &  -i q^{\prime}_{\nu} \Lambda^{\mu a;\nu b}_{AV}(p',q',p,q) = i \epsilon_{bac} \Lambda^{\mu c}_{A}(p^{\prime},p) \nonumber\\
& &  -\Lambda^{\mu b}_{A}(p+q,p) \frac{\tau_a}{2} - \frac{\tau_a}{2} \Lambda^{\mu b}_{A}(p^{\prime},p^{\prime}-q). \label{LAVVWI}
\end{eqnarray}

A solution that fulfills Eq.(\ref{LAVVWI}) and Eq.(\ref{LAVAWI}) in the chiral limit is given by:
\begin{eqnarray}
& & \Lambda^{\mu a;\nu b}_{AV}(p',q',p,q) = \intw \Big\{ \frac{i}{\slash{p}'-\omega} \gamma^{\nu} \frac{\tau^b}{2} \frac{i}{\slash{p}+\slash{q}-\omega} \nonumber \\
& & \times \Big(  g_A\gamma^{\mu} - 2\omega\frac{q^{\mu}}{q^2} + (1-g_A)\frac{q^{\mu}\slash{q}}{q^2} \Big) \gamma^5 \frac{\tau_a}{2} \frac{i}{\slash{p}-\omega} \nonumber \\
& & + \frac{i}{\slash{p}'-\omega} \Big(  g_A\gamma^{\mu} - 2\omega\frac{q^{\mu}}{q^2} + (1-g_A)\frac{q^{\mu}\slash{q}}{q^2} \Big) \gamma^5 \frac{\tau_a}{2} \nonumber \\
& & \times \frac{i}{\slash{p}'-\slash{q}-\omega} \gamma^{\nu} \frac{\tau^b}{2} \frac{i}{\slash{p}-\omega} \Big\}, \label{LAV}
\end{eqnarray}
with $p'+q'=p+q$. In fact, replacing Eq.(\ref{LAV}) in Eq.(\ref{LAVAWI}) one obtains:
\begin{eqnarray}
& & -i q_{\mu} \Lambda^{\mu a;\nu b}_{AV}(p',q',p,q) = i \epsilon_{bac} \Lambda^{\nu c}_{A}(p^{\prime},p)\Big|_{g_A=1} \nonumber \\
& & + \Lambda^{\nu b}_{V}(p^{\prime},p+q)\gamma^5 \frac{\tau_a}{2} + \gamma^5\frac{\tau_a}{2} \Lambda^{\nu b}_{V}(p^{\prime}-q,p) \nonumber \\
& & + i 2 \epsilon_{bac} \frac{q^{\nu}}{q^2} \intw \omega \frac{i}{\slash{p}'-\omega} \gamma^5 \frac{\tau^c}{2} \frac{i}{\slash{p}-\omega} \label{PCAC}
\end{eqnarray}

Eq.(\ref{PCAC}) displays the expected partial conservation of the axial current. This can be verified by evaluating the last term in Eq.(\ref{PCAC}) 
with $\slash{p}'=\slash{p}$, resulting in
\begin{eqnarray}
& &  \Big\{ 2i \epsilon_{bac} \frac{q^{\nu}}{q^2} \intw \omega \frac{i}{\slash{p}'-\omega} \gamma^5 \frac{\tau^c}{2} \frac{i}{\slash{p}-\omega} \Big\}_{\slash{p}'=\slash{p}} \nonumber \\
& &  = \Big\{ 2i \epsilon_{bac} \frac{q^{\nu}}{q^2} \intw \omega \frac{\slash{p}'\slash{p}-\omega^2 - \omega(\slash{p}'-\slash{p})}{(p'^2-\omega^2)(p^2-\omega^2)} \gamma^5 \frac{\tau^c}{2}\Big\}_{\slash{p}'=\slash{p}} \nonumber \\
& &  =  2i \epsilon_{bac} \frac{q^{\nu}}{q^2}  \gamma^5 \frac{\tau^c}{2} \intw \frac{\omega}{(p^2-\omega^2)} \nonumber \\
& &  =  2i \epsilon_{bac} m(p) \frac{q^{\nu}}{q^2} \gamma^5 \frac{\tau^c}{2} \intw \frac{1}{(p^2-\omega^2)}  \label{termPCAC}
\end{eqnarray}
where we have made use of Eq.(\ref{massfunction}). So, the last term of Eq.(\ref{PCAC}) shows the violation of the axial current conservation
arising from the fermionic mass term, as usual. When chiral symmetry is restored, Eq.(\ref{termPCAC}) will vanish, and the axial Ward identity,
Eq.(\ref{LAVAWI}), will be fulfilled by the solution (\ref{LAV}). 

In the SQM the probability amplitude for the axial-vector-vector process is obtained by closing the quark line in the unamputated two currents (axial and vector) vertex, Eq.(\ref{LAV}). 
With the appropriated insertion of the charge matrices, and the momenta labels chosen as in Fig.(\ref{fig1}), this probability amplitude is given by:
\begin{eqnarray}
& & T^{AVV}_{\rho \mu \nu}(q_1,q_2) = - \intw \int_p Tr \bigg[ N_{C}i \tau^{3} \nonumber \\
& & \Big( g_A(\omega) \gamma_{\rho} - \frac{2\omega}{q^2} q_{\rho} + \frac{1-g_A(\omega)}{q^2} q_{\rho} {\slash q} \Big) \gamma_{5} \frac{i}{{\slash p} + {\slash q}_1 - \omega}  \nonumber \\
& & i \hat{Q} \gamma_{\mu} \frac{i}{{\slash p} - \omega} i \hat{Q} \gamma_{\nu} \frac{i}{{\slash p} - {\slash q}_2 - \omega} \bigg] + \text{crossed terms}, \label{TAVV}
\end{eqnarray}
where $\hat{Q}$ is the quark charge matrix, $q=q_1+q_2$ and, for a $SU(2)$ model, $\tau^3$ is the Pauli matrix $\sigma_z$. 
In order to compare with  the PCAC relation, let us 
rewrite the term proportional to $\frac{\omega}{q^2} q_{\rho}$ as
\begin{eqnarray}
&&T_{\mu \nu}=\frac{1}{m} \intw \omega \int_p Tr \bigg\{ \frac{\gamma_5 ({\slash p} + {\slash q}_1 + \omega)}{((p+q_1)^2-\omega^2)} \times \nonumber\\
&&\frac{\gamma_{\mu}({\slash p} + \omega)\gamma_{\nu}({\slash p} - {\slash q}_2 + \omega)}{(p^2-\omega^2)((p-q_2)^2-\omega^2)}  \bigg\} + \text{crossed terms} \label{Tmunu}
\end{eqnarray}
where we have introduced the mass function at zero external momentum $m$ in order to associate $T_{\mu \nu}$
with the standard representation of the neutral pion to two photons decay amplitude. Eq.(\ref{TAVV}), of course, does not depend on $m$, 
and can be rewritten as
\begin{equation}
T^{AVV}_{\rho \mu \nu}(q_1,q_2) = \lambda  \Big(\hat{T}_{\rho \mu \nu} + \frac{q_{\rho}}{q^2} 2 m T_{\mu \nu} + \frac{q_{\rho}}{q^2} T^{*}_{\mu \nu} \Big), \label{TAVV2}
\end{equation}
with
\begin{eqnarray}
&&\hat{T}_{\rho \mu \nu}=\intw g_A(\omega) \int_p Tr \bigg\{ \frac{\gamma_{\rho}\gamma_5 ({\slash p} + {\slash q}_1 + \omega)}{((p+q_1)^2-\omega^2)} \times \nonumber\\
&&\frac{\gamma_{\mu}({\slash p} + \omega)\gamma_{\nu}({\slash p} - {\slash q}_2 + \omega)}{(p^2-\omega^2)((p-q_2)^2-\omega^2)}  \bigg\} + \text{crossed terms} \label{That}
\end{eqnarray}
and
\begin{eqnarray}
&&T^{*}_{\mu \nu}=\intw (1-g_A(\omega)) \int_p Tr \bigg\{ \frac{{\slash q}\gamma_5 ({\slash p} + {\slash q}_1 + \omega)}{((p+q_1)^2-\omega^2)} \times \nonumber\\
&&\frac{\gamma_{\mu}({\slash p} + \omega)\gamma_{\nu}({\slash p} - {\slash q}_2 + \omega)}{(p^2-\omega^2)((p-q_2)^2-\omega^2)}  \bigg\} + \text{crossed terms}. \label{Tstar}
\end{eqnarray}
We also have
\begin{equation}
\lambda = N_c Tr\{\tau ^{3}\hat{Q}\hat{Q}\}=1.
\end{equation}

It is interesting to note that, after taking the Dirac traces, Eq.(\ref{Tmunu}) is logarithmically divergent, and thus
does not present surface terms in its computation. Eq.(\ref{Tstar}) is quadratically divergent, and in principle should
present surface terms. However, in the sharp cutoff regularization scheme these surface terms result zero. After a little algebra,
we obtain
\begin{equation}
T^{*}_{\mu \nu}= \frac{1}{2 \pi^2} \epsilon_{\rho \mu \nu \eta} q_1^{\rho}q_2^{\eta}\intw (1-g_A(\omega)), \label{Tstar2}
\end{equation}
and
\begin{equation}
2 m T_{\mu\nu}=\frac{2}{4\pi^2}\epsilon_{\mu\nu\rho\eta}q_1^{\eta}q_2^{\rho} \intw. 
\end{equation}

Eq.(\ref{That}), however, presents non-null surface terms in its calculation. These surface terms come from the difference of
logarithmically divergent integrals, the same integrals in $p$ appearing in Eq.(\ref{surfacePiPS}),
\begin{equation}
\int_p \Big( \frac{g^{\mu\nu}}{(p^2-\omega^2)^2}-4\frac{p^{\mu}p^{\nu}}{(p^2-\omega^2)^3} \Big).
\end{equation}
As already discussed,
this term corresponds to a regularization ambiguity, since it can result zero in some regularizations (e.g. gauge invariant Pauli-Villars)
or finite in other ones ($=\frac{i}{2(4\pi^2)}$ in the sharp cutoff regularization) \footnote{In addition to the surface terms coming from the internal momentum shift, 
there are also regularization ambiguities that appear due to the tensorial character of the amplitudes, as the vacuum polarization 
tensor in QED is a classical example \cite{Battistel:1998sz}. Here, we are not isolating this second type of ambiguity.}.
So, introducing two Feynman parameters on Eq.(\ref{That}) we obtain
\begin{eqnarray}
&& \hat{T}_{\rho \mu \nu}=\intw g_A(\omega) \int_0^1 dy \int_0^{1-y} dx \label{feynpar} \\
&& \times \int_p \frac{Tr\{ \gamma_\rho \gamma_5 (\slash{p}+\slash{q}_1+\omega) \gamma_{\mu} (\slash{p}+\omega) \gamma_{\nu} (\slash{p}-\slash{q}_2+\omega) \}}{((p+q_1 x-q_2 y)^2-M^2(\omega,x,y))^3}, \nonumber
\end{eqnarray}
with $M^2(\omega,x,y)=q_1^2x(x-1)+q_2^2y(y-1)-2q_1q_2xy+\omega^2$ (the sharp cut-off regularization scheme is, as before, implicitly assumed).
In several regularization schemes, as in the sharp-cutoff,
we are not allowed to shift the $p$ variable in Eq.(\ref{feynpar}), unless we introduce the corresponding surface term. Following
the procedure employed in Ref. \cite{Battistel:1998sz}, we obtain 
\begin{eqnarray}
&& \hat{T}_{\rho \mu \nu}=S_{\rho \mu \nu} + \intw g_A(\omega) \int_0^1 dy \int_0^{1-y} dx \nonumber \\
&& \times \int_p \frac{1}{(p^2-M^2(\omega,x,y))^3} \nonumber \\
&& \times Tr\{ \gamma_\rho \gamma_5 (\slash{p}+(1-x)\slash{q}_1+y\slash{q}_2+\omega) \gamma_{\mu}(\slash{p}-x\slash{q}_1 \nonumber \\
&&  +y\slash{q}_2+\omega) \gamma_{\nu} (\slash{p}-x\slash{q}_1-(1-y)\slash{q}_2+\omega) \}, 
\end{eqnarray}
where $S_{\rho \mu \nu}$ is the surface term given by
\begin{eqnarray}
&& S_{\rho \mu \nu} = \intw g_A(\omega) \int_0^1 dy \int_0^{1-y} dx \nonumber \\
&& \times \int_p (q_1^{\sigma}x-q_2^{\sigma}y) \frac{\partial}{\partial p^{\sigma}} \Big\{ \frac{Tr\{ \gamma_\rho \gamma_5 \slash{p} \gamma_{\mu} \slash{p} \gamma_{\nu} \slash{p} \}}{(p^2-M^2(\omega,x,y))^3} \Big\}
\end{eqnarray}
 
Of course, as usual, $T^{AVV}_{\rho \mu \nu}$ is finite, so all integrals can be computed and the connection limit $\Lambda \rightarrow \infty$
can be taken. After doing that, we have
\begin{eqnarray}
&&\hat{T}_{\rho \mu \nu}= \frac{1}{4 \pi^2} \bigg\{ \int d\omega \frac{g_A(\omega)\rho(\omega)}{6\omega^2} q_1^{\eta}q_2^{\beta}\{ \epsilon_{\rho\mu\beta\eta}(q_{2\nu}+2q_{1\nu}) \nonumber \\
&&+ \epsilon_{\rho\beta\nu\eta}(2q_{2\mu}+q_{1\mu}) \} \nonumber \\
&&+ \frac{2}{3} \epsilon_{\rho\mu\nu\eta} \intw g_A(\omega) (q_2^{\eta}-q_1^{\eta}) \bigg\} + S_{\rho \mu \nu}  \label{That2}
\end{eqnarray}
with
\begin{eqnarray}
&&S_{\rho \mu \nu}= \frac{1}{12\pi^2} \epsilon_{\rho\mu\nu\eta} \intw g_A(\omega) (q_2^{\eta}-q_1^{\eta}) \label{surfterm}
\end{eqnarray}

From Eqs. (\ref{TAVV2}), (\ref{That2}) and (\ref{Tstar2}), the evaluation of the Ward-Takahashi identities in momentum space results in
\begin{eqnarray}
&& q_1^{\mu}T^{AVV}_{\rho \mu \nu} = \frac{1}{4\pi^2} \frac{2}{3} q_1^{\mu}q_2^{\eta}\epsilon_{\mu\nu\rho\eta}\intw g_A(\omega) \nonumber \\
&& + q_1^{\mu}S_{\rho \mu \nu}, \label{EMWTI}
\end{eqnarray}
for the VWI, with a similar expression for $q_2^{\nu}T^{AVV}_{\rho \mu \nu}$, and
\begin{eqnarray}
&&(q_1^{\rho}+q_2^{\rho})T^{AVV}_{\rho \mu \nu} = \frac{2}{4\pi^2} q_1^{\eta}q_2^{\rho}\epsilon_{\mu\nu\rho\eta} \frac{1}{3}\intw g_A(\omega) \nonumber \\
&& + (q_1^{\rho}+q_2^{\rho}) S_{\rho \mu \nu} + 2mT_{\mu\nu}- \frac{2}{4\pi^2} q_1^{\eta}q_2^{\rho}\epsilon_{\mu\nu\rho\eta}, \label{CWTI}
\end{eqnarray}
for the AWI.

Before performing the spectral mass integrals, the result is potentially ambiguous due the presence of the surface term
$S_{\rho \mu \nu}$ in Eqs.(\ref{That2}), (\ref{EMWTI}) and (\ref{CWTI}). 
For a constant ($\omega$ independent) $g_A$, the surface term, Eq.(\ref{surfterm}), depends on the choice of the intermediary regularization employed
in the evaluation of Eq.(\ref{TAVV}), and it could generate different results in the Ward identity to be violated by the anomaly, as it is well known.

Nevertheless,  from Eq.(\ref{surfterm}) we can see that the computation of the axial-vector-vector amplitude in the SQM with
the axial vertex given by Eq.(\ref{AxialVertex2}) and with $g_A=\alpha \omega$ (or $g_A=\alpha \omega^2$) is free from surface terms - the spectral condition, Eq.(\ref{positivemoments}),
ensures their vanishing, as well as the vanishing of the other terms proportional to $g_A(\omega)$ in Eqs.(\ref{EMWTI})
and (\ref{CWTI}). 
In this case, the VWI is always preserved (i.e., is preserved in an ambiguity free way) and the AWI
 is violated. Hence, we clearly see that the choice of the specific dependence of the axial coupling with
the spectral mass can result in an ambiguity free result, with the conservation of the VWI, and
the violation of the chiral one, as expected for QCD. In this case, the vanishing of the spectral axial coupling 
in the zero spectral mass limit does not
imply in the vanishing of the axial coupling itself. However, it is also possible to obtain the violation
of the VWI, with conservation of the axial current, when the spectral axial coupling depends on the spectral mass
with a power lower than 1. In this case, the presence of ambiguous terms implies in the freedom on choosing which
Ward identity is to be violated. 

\section{Conclusions}
We have investigated the chiral anomaly in the context of the Spectral Quark Model. We have proposed a generalized form of
the four points vertex function with one axial and one vector current which includes the possibility of a non unitary,
spectral mass dependent, axial coupling. This vertex function displays the expected partial conservation of the axial current,
with the chiral Ward identity being violated by a term proportional to the mass function which vanishes at the chiral limit.
The triangle anomaly was computed, taking into account the surface term that appears in its calculation, and we have shown 
that a dependence of the axial coupling on integer positive powers of the spectral mass is necessary in order to render
the triangle amplitude free from ambiguities. In this case, the vector Ward identity is preserved, with the chiral
Ward identity being violated as expected for QCD. We also remarked that the dependence of the pion to quarks couplings
on the spectral mass, via the Goldberger-Treiman relation for the pseudoscalar coupling, and via the ansatz employed here
for the axial coupling, are essential to the obtaining of a result free from divergences and regularization ambiguities.

In summary, we have shown that, in the context of the Spectral Quark Model, the chiral anomaly computation can be carried out 
reproducing the expected result without any ambiguity introduced by regularizations schemes, if the
axial vertex is treated as a non-local spectral mass dependent vertex. This shows to be compatible with an non-unitary
axial vector coupling, not vanishing at the chiral limit. Our result suggests that the Spectral Quark Model provides an useful mechanism to justify why in QCD the anomaly violates the chiral 
symmetry, instead of violating gauge symmetry, as in the case of the non conservation of the barionic number.
However, we also discussed that the more general form for the axial vertex leaves room to the violation
of the vector Ward Identity, complying the axial one.
It could be interesting to analyze these features of the Spectral Quark Model in the QCD chiral phase transition 
\cite{Chandrasekharan:2006zq,Chandrasekharan:2007up} or in the construction of spectral approaches to some recent 
applications of the chiral anomaly \cite{Bracken:2008zza,Benfatto:2004ib,Sasaki:2001zb,Klebanov:2002gr}.

\section{Acknowledgments}
This research was supported by CAPES-Brazil. The authors would like to thank H. Caldas for carefully reading the manuscript.


\end{document}